\documentclass[12pt]{article}
\usepackage{graphicx}
\usepackage{lscape}
\usepackage{citesort}
\usepackage{amssymb}
\usepackage{appendix}
\usepackage{multirow}

\begin{document}
\def\qq{\langle \bar q q \rangle}
\def\uu{\langle \bar u u \rangle}
\def\dd{\langle \bar d d \rangle}
\def\sp{\langle \bar s s \rangle}
\def\GG{\langle g_s^2 G^2 \rangle}
\def\Tr{\mbox{Tr}}
\def\figt#1#2#3{
        \begin{figure}
        $\left. \right.$
        \vspace*{-2cm}
        \begin{center}
        \includegraphics[width=10cm]{#1}
        \end{center}
        \vspace*{-0.2cm}
        \caption{#3}
        \label{#2}
        \end{figure}
    }

\def\figb#1#2#3{
        \begin{figure}
        $\left. \right.$
        \vspace*{-1cm}
        \begin{center}
        \includegraphics[width=10cm]{#1}
        \end{center}
        \vspace*{-0.2cm}
        \caption{#3}
        \label{#2}
        \end{figure}
                }

\def\ds{\displaystyle}
\def\beq{\begin{equation}}
\def\eeq{\end{equation}}
\def\bea{\begin{eqnarray}}
\def\eea{\end{eqnarray}}
\def\beeq{\begin{eqnarray}}
\def\eeeq{\end{eqnarray}}
\def\ve{\vert}
\def\vel{\left|}
\def\ver{\right|}
\def\nnb{\nonumber}
\def\ga{\left(}
\def\dr{\right)}
\def\aga{\left\{}
\def\adr{\right\}}
\def\lla{\left<}
\def\rra{\right>}
\def\rar{\rightarrow}
\def\lrar{\leftrightarrow}
\def\nnb{\nonumber}
\def\la{\langle}
\def\ra{\rangle}
\def\ba{\begin{array}}
\def\ea{\end{array}}
\def\tr{\mbox{Tr}}
\def\ssp{{\Sigma^{*+}}}
\def\sso{{\Sigma^{*0}}}
\def\ssm{{\Sigma^{*-}}}
\def\xis0{{\Xi^{*0}}}
\def\xism{{\Xi^{*-}}}
\def\qs{\la \bar s s \ra}
\def\qu{\la \bar u u \ra}
\def\qd{\la \bar d d \ra}
\def\qq{\la \bar q q \ra}
\def\gGgG{\la g^2 G^2 \ra}
\def\q{\gamma_5 \not\!q}
\def\x{\gamma_5 \not\!x}
\def\g5{\gamma_5}
\def\sb{S_Q^{cf}}
\def\sd{S_d^{be}}
\def\su{S_u^{ad}}
\def\sbp{{S}_Q^{'cf}}
\def\sdp{{S}_d^{'be}}
\def\sup{{S}_u^{'ad}}
\def\ssp{{S}_s^{'??}}

\def\sig{\sigma_{\mu \nu} \gamma_5 p^\mu q^\nu}
\def\fo{f_0(\frac{s_0}{M^2})}
\def\ffi{f_1(\frac{s_0}{M^2})}
\def\fii{f_2(\frac{s_0}{M^2})}
\def\O{{\cal O}}
\def\sl{{\Sigma^0 \Lambda}}
\def\es{\!\!\! &=& \!\!\!}
\def\ap{\!\!\! &\approx& \!\!\!}
\def\md{\!\!\!\! &\mid& \!\!\!\!}
\def\ar{&+& \!\!\!}
\def\ek{&-& \!\!\!}
\def\kek{\!\!\!&-& \!\!\!}
\def\cp{&\times& \!\!\!}
\def\se{\!\!\! &\simeq& \!\!\!}
\def\eqv{&\equiv& \!\!\!}
\def\kpm{&\pm& \!\!\!}
\def\kmp{&\mp& \!\!\!}
\def\mcdot{\!\cdot\!}
\def\erar{&\rightarrow&}

% .........................................................

\def\simlt{\stackrel{<}{{}_\sim}}
\def\simgt{\stackrel{>}{{}_\sim}}

% .........................................................

\title{
         {\Large
                 {\bf
                     Form factors for the rare $\Lambda_{b}(\Lambda_{b}^*)\rar
                     N\ell^+\ell^- $ decays in light cone QCD sum rules
                 }
         }
      }

\author{\vspace{1cm}\\
{\small T. M. Aliev\,$^a\!\!$ \thanks {e-mail:
taliev@metu.edu.tr}\,\,,
T. Barakat\,$^b\!\!$ \thanks {e-mail:
tbarakat@ksu.edu.sa}\,\,, M. Savc{\i}\,$^a\!\!$ \thanks
{e-mail: savci@metu.edu.tr}} \\
{\small $^a$ Physics Department, Middle East Technical University,
06531 Ankara, Turkey} \\
{\small $^b$ Physics Department, King Saud University} \\
{\small Riyadh 11451, Saudi Arabia} }
\date{}

\begin{titlepage}
\maketitle
\thispagestyle{empty}

\begin{abstract}
\baselineskip 0.5 cm
Form factors of the rare $\Lambda_{b}(\Lambda_{b}^*)\to N\ell^{+}\ell^{-}$ 
decays are calculated in the framework of the light cone QCD sum rules
by taking into account of the contributions from the negative parity
baryons. Using the obtained results on the form factors,
the branching ratios of the considered decays are estimated.
The numerical survey for the branching ratios of the
$\Lambda_b \rar N\ell^+\ell^- $ and $\Lambda_b^\ast \rar N\ell^+\ell^- $
decays indicate that these transitions could be measurable in LHCb in near
future. Comparison of our predictions on the form factors and branching
ratios with those  existing in the literature is also performed.
\\
\vspace{0.5cm} PACS numbers: 12.15.Ji; 12.60.-i\\
\\
\end{abstract}
\end{titlepage}

\section{Introduction}

Lately, impressive experimental progress has been made in investigation
of the rare decays of heavy $\Lambda_{b}$ baryon. The CDF collaboration
\cite{Rbrk01} announced the first evidence of the rare $\Lambda_{b}\to
\Lambda \mu^{+}\mu^{-}$ decay \cite{Rbrk02}. Very recently, the suppressed
$\Lambda_{b}\to p\pi^{-}\mu^{-}\mu^{+}$ decay, excluding $J/\psi$ and
$\psi(2S)\to \mu^{+}\mu^{-}$ resonance is observed \cite{Rbrk03}. This is
the first observation of  a $b\to d$ transition in a baryonic decay.
The measured branching ratio is {\cal B}($\Lambda_{b}\to p\pi^{-}\mu^{+}
\mu^{-}$) =$ \left(6.9\pm 1.9 \pm 1.1_{-1.0}^{+1.3}\right)\times 10^{-8}$.
Next, the LHCb Collaboration performed a detailed analysis of differential
branching ratio, forward--backward angular distributions, and asymmetries in
the meson and hadronic systems \cite{Rbrk04}. This evidence stimulated the
search of other similar decays, such as $\Lambda_{b}\to N\mu^{-}\mu^{+}$,
which can in principle be discovered in the near future at LHCb.

The detailed study of $\Lambda_{b}$ baryon decays receives special attention
for two reasons. The first reason is due to the fact that the $\Lambda_{b}$
baryon has spin one-half and therefore can give essential information about
the helicity structure of the effective Hamiltonian. The second reason is that,
many aspects of the effective theory can be tested.

In the present work, we study the $\Lambda_{b}\to N\mu^{-}\mu^{+}$ decay within
light cone QCD sum rules (LCSR) \cite{Rbrk05}. This method was applied to study
the electromagnetic form factors of the nucleon \cite{Rbrk06,Rbrk07}, and
$\Sigma$, $\Lambda$ baryons \cite{Rbrk08,Rbrk09}, as well as to the study of
the rare $\Lambda_{b}\to \Lambda\ell^{+}\ell^{-}$ transition \cite{Rbrk10}.
This rare decay has comprehensively been studied in framework of the
different approaches such as, relativistic quark model \cite{Rbrk11},
lattice QCD \cite{Rbrk12}, soft collinear effective theory \cite{Rbrk13},
heavy quark effective theory \cite{Rbrk14}, and etc.

The plan of this work is organized as follows. In Section 2, the
LCSR for the $\Lambda_{b}(\Lambda_b^\ast)\to N$ 
form factors are derived. The numerical analysis
of the LCSR for the form factors obtained in the previous section is presented in
Section 3. Using these form factors, in this section we also calculate the decay
widths $\Lambda_{b}(\Lambda_b^\ast)\to N\ell^{+}\ell^{-}$ $(\ell =
 e, \mu, \tau)$ transitions.

\section{Transition form factors for the $\Lambda_b(\Lambda_b^\ast)\to N
\ell^{+}\ell^{-}$ decays in light cone sum rules}              

At the quark level, $\Lambda_b(\Lambda_b^\ast)\to N \ell^{+}\ell^{-}$ 
decay is governed by the flavor changing neutral current $b\to d$ transition.
The hadronic matrix elements responsible for $\Lambda_{b}(\Lambda_b^\ast)\to N$
 transition are determined by considering the transition current between 
$\Lambda_{b}(\Lambda_b^\ast)$ and $N$ states. The relevant form factors
of the vector, axial-vector, and tensor currents are defined as,
\bea
\label{ebrk01}
\langle \Lambda_{Q}(p-q) \md  \bar b \gamma_\mu (1-\gamma_5) d \mid N(p)\rangle =
\bar {u}_\Lambda(p-q) \Big[f_{1}(q^{2})\gamma_{\mu}+{i} \frac{f_{2}(q^{2})}
{m_{\Lambda_b}}\sigma_{\mu\nu}q^{\nu} + \frac{f_{3}(q^{2})}{m_{\Lambda_b}} q^{\mu} \nnb \\
\ek g_{1}\gamma_{\mu}\gamma_5 (q^{2})-{i} \frac{g_{2}(q^{2})}{m_{\Lambda_b}}
\sigma_{\mu\nu}\gamma_5q^{\nu}-\frac{g_{3}(q^{2})}{m_{\Lambda_b}} q^{\mu}\gamma_5
\vphantom{\int_0^{x_2}}\Big] u_{N}(p)~,
\eea
and
\bea
\label{ebrk02}
\langle \Lambda(p-q)\md \bar b i \sigma_{\mu\nu}q^{\nu} (1+ \gamma_5)
d \mid N(p)\rangle =\bar{u}_\Lambda(p-q)
\Big[\frac{f_{1}^{T}(q^{2})}{m_{\Lambda_b}}(\gamma_{\mu}q^2-q_{\mu}\not\!q) \nnb \\ 
\ar {i}f_{2}^{T}(q^{2})\sigma_{\mu\nu}q^{\nu}
+  \frac{g_{1}^{T}(q^{2})}{m_{\Lambda_b}}(\gamma_{\mu}q^2-q_{\mu}\not\!q)\gamma_5
+{i}g_{2}^{T}(q^{2})\sigma_{\mu\nu}\gamma_5 q^{\nu}\vphantom{\int_0^{x_2}}\Big] u_{N}(p)~,
\eea

In order to determine the form factors $f_i,~f_i^T$ in Eqs. (\ref{ebrk01})
and (\ref{ebrk02})
we introduce the following correlation functions,
\bea
\label{ebrk03}
\Pi^{I}_{\alpha}(p,q) =
i\int d^{4}xe^{iqx}\langle 0 \mid T\{\eta_{\Lambda_{b}}(0) J_{\alpha}(x)\}\mid
N(p)\rangle~,
\eea
where $\eta_{\Lambda_{b}}$ is the interpolating current of $\Lambda_{b}$-baryon,
$ J_{\alpha}(x)$ is the heavy--light transition current, and $\alpha$ index
corresponds to the choice of form of interpolating current. For the decays
at hand these currents are  $J_{\alpha}$=  $\bar b \gamma_\mu(1-\gamma_5) d$
or $\bar{b}~i\sigma_{\mu\nu}q^{\nu} (1+ \gamma_5)d$.

The most general form of the interpolating current for the $\Lambda_b$
baryon is given as,
\bea
\label{ebrk04}
\eta_{\Lambda_{b}}\es\frac{1}{\sqrt{6}}\epsilon_{abc}
\Bigg\{\vphantom{\int_0^{x_2}}2\Big[(u^{aT}(x)Cd^{b}(x))\gamma_{5}b^{c}(x)
+\beta(u^{aT}(x)C\gamma_{5}d^{b}(x))b^{c}(x)\Big] \nnb \\
\ar (u^{aT}(x)Cb^{b}(x))\gamma_{5}d^{c}(x)
+\vphantom{\int_0^{x_2}}
\beta(u^{aT}(x)C\gamma_{5}b^{b}(x))d^{c}(x) \nnb \\
\ar (b^{aT}(x)Cd^{b}(x))
\gamma_{5}u^{c}(x)
+\beta(b^{aT}(x)C\gamma_{5}d^{b}(x))u^{c}(x)\Bigg\}~,
\eea
where $a,~b$ and $c$ are the color indices, $C$ is the charge conjugation
operator, and $\beta$ is an arbitrary parameter with $\beta=-1$
corresponding to the Ioffe current.

The first step in deriving the sum rules for the transition form factors
is to insert the full set of beauty-baryon states between the
interpolating current $\eta_{\Lambda_{b}}$, and the transition current $J_{\alpha}$,
and then the ground state contributions are isolated. At this point we meet the
following problem which is absent for the mesonic system. The interpolating
current of the baryon has nonzero overlap not only with the ground state with
$J^{P}={\frac{1}{2}}^+$, but also with the negative parity $J^{P}={\frac{1}{2}}^-$
baryon. Calculations show that the
mass difference between the negative and positive parity states of the heavy 
$\Lambda_{b}$ baryon is about $(250-300)~ MeV$ \cite{Rbrk15}.
Therefore negative parity baryons can give considerable contribution
to the sum rules. For this reason the standard quark--hadron duality approximation
should be modified, which leads to the strong dependence of the sum rules
predictions on the choice of interpolating current. Keeping these preliminary
remarks in mind, the expression for the physical part of the correlation
can be written as,
\bea
\label{ebrk05}
\Pi_{\mu}^{I}(p,q)=\sum_{i=+,-}\frac{\langle 0\mid \eta_{\Lambda_{b}}(0)
\mid \Lambda_{b}^{i}(p-q,s)\rangle\langle
\Lambda_{b}^{i}(p-q,s)\mid  \bar b \gamma_\mu (1-\gamma_5) d \mid
N(p)\rangle}{m_{i}^{2}-(p-q)^{2}}~,
\eea

\bea
\label{ebrk06}
\Pi_{\mu}^{II}(p,q)=\Pi_{\mu}^{I}(p,q)\Big[\gamma_\mu (1-\gamma_5)\rightarrow 
i \sigma_{\mu\nu}q^{\nu} (1+ \gamma_5)\Big]~,
\eea
where summation is performed over the positive and negative parity
$\Lambda_b$ baryons.
The first matrix element in Eqs. (\ref{ebrk05} and (\ref{ebrk06}) is
defined in terms of the residues of $\Lambda_b$ and $\Lambda_b^\ast$ 
baryons as follows
\bea
\label{ebrk07}
\langle 0 \mid  \eta_{\Lambda_b} \mid
\Lambda_b (p-q)\rangle = \lambda_{\Lambda_b} u_{\Lambda_b}(p-q)~,
\eea
\bea
\label{ebrk08}
\langle 0 \mid  \eta_{\Lambda_b^\ast} \mid
\Lambda_b^\ast(p-q)\rangle=\lambda_{\Lambda_b^\ast}\gamma_5
u_{\Lambda_b^\ast}(p-q)~.
\eea

Using Eqs. (\ref{ebrk01}), (\ref{ebrk02}), (\ref{ebrk07}),
and (\ref{ebrk08}), the hadronic part of the correlation function can be
written as,
\bea
\label{ebrk09}
\Pi_{\mu}^{I}(p,q) \es \frac{\lambda_{\Lambda_b}}{m_{\Lambda_b}^{2}-(p-q)^{2}}\Bigg\{ f_{1}(q^{2})
( 2p_{\mu}-m_N\gamma_{\mu}-2q_{\mu}+\gamma_{\mu}\not\!q + m_{\Lambda_b}\gamma_{\mu} ) \nnb\\
\ek \frac{f_{2}(q^{2})}{m_{\Lambda_b}} \Big[2p_{\mu}\not\!q +\gamma_{\mu}(m_{\Lambda_b}^2-m_N^2)+
(m_{\Lambda_b}+m_N)\gamma_{\mu}\not\!q
-(m_{\Lambda_b}+m_N)q_{\mu}-q_{\mu}\not\!q\Big] \nnb\\
\ar f_{3}(q^{2})\frac{q_{\mu}}{m_{\Lambda_b}} ( m_{\Lambda_b}+ m_N-\not\!q )-g_{1}(q^{2})
\Big[ 2p_{\mu}\gamma_5+(m_{\Lambda_b}+ m_N)\gamma_{\mu}\gamma_5 -2q_{\mu}\gamma_5
\nnb \\
\ar \gamma_{\mu} \not\!q\gamma_5 \Big]
+ \frac{g_{2}(q^{2})}{m_{\Lambda_b}}\Big[ 2p_{\mu}\not\!q\gamma_5+\gamma_{\mu}
\gamma_5(m_{\Lambda_b}^2-m_N^2) +(m_{\Lambda_b}-m_N)\gamma_{\mu}\not\!q\gamma_5
\nnb \\ 
\ek(m_{\Lambda_b}-m_N) q_{\mu}\gamma_5-q_{\mu}\not\!q\gamma_5 \Big]
- g_{3}(q^{2})\frac{q_{\mu}}{m_{\Lambda_b}} \Big[ (m_{\Lambda_b}-m_N)\gamma_5-
\not\!q \gamma_5\Big]\Bigg\} \nnb\\
\ar\frac{\lambda_{\Lambda_b^\ast}}{m_{\Lambda_b^\ast}^{2}-(p-q)^{2}}
\Bigg\{\widetilde{f}_1(q^{2})
(- 2 p_{\mu}+m_N\gamma_{\mu}+2q_{\mu}-\gamma_{\mu}\not\!q + 
m_{\Lambda_b^\ast}\gamma_{\mu}) \nnb\\
\ar \frac{\widetilde{f}_2(q^{2})}{m_{\Lambda_b^\ast}} \Big[2p_{\mu}\not\!q +\gamma_{\mu}
(m_{\Lambda_b^\ast}^2-m_N^2)-(m_{\Lambda_b^\ast}-m_N)\gamma_{\mu}\not\!q +
(m_{\Lambda_b^\ast}-m_N)q_{\mu}-q_{\mu} \not\!q\Big] \nnb\\
\ar \widetilde{f}_3(q^{2})\frac{q_{\mu}}{m_{\Lambda_b^\ast}} (
m_{\Lambda_b^\ast}-m_N+\not\!q )-
\widetilde{g}_1(q^{2})\Big[-2p_{\mu}\gamma_5+(m_{\Lambda_b^\ast}-m_N)\gamma_{\mu}\gamma_5
+ 2q_{\mu}\gamma_5\nnb \\
\ek\gamma_{\mu}\not\!q\gamma_5 \Big]
+ \frac{\widetilde{g}_2(q^{2})}{m_{\Lambda_b^\ast}}\Big[
-2p_{\mu}\not\!q\gamma_5-
\gamma_{\mu}\gamma_5(m_{\Lambda_b^\ast}^2-m_N^2)
+(m_{\Lambda_b^\ast}+m_N)\gamma_{\mu}\not\!q\gamma_5 \nnb \\
\ek (m_{\Lambda_b^\ast}+m_N)q_{\mu}\gamma_5+q_{\mu}\not\!q\gamma_5 \Big]
- \widetilde{g}_3(q^{2})\frac{q_{\mu}}{m_{\Lambda_b^\ast}} \Big[(m_{\Lambda_b^\ast}+m_N)\gamma_5
+\not\!q\gamma_5\Big]\Bigg\}~,\\ \nnb \\
%%%%%%%%%
\label{ebrk10}
\Pi_{\mu}^{II}(p,q) \es \frac{\lambda_{\Lambda_b}}{m_{\Lambda_b}^{2}-(p-q)^{2}}\Bigg\{
{f_1^T(q^{2})\over m_{\Lambda_b}}
\Big[ \Big((m_{\Lambda_b}- m_N)\gamma_\mu + \gamma_\mu\not\!q + 2p_\mu\Big) q^2 \nnb \\
\ek \Big((m_{\Lambda_b}- m_N) \not\!q - (m_{\Lambda_b}^2- m_N^2 -
2 q^2)\Big)q_\mu\Big] + f_2^T \Big[ (-2 p_\mu +q_\mu)\not\!q  \nnb \\
\ek (m_{\Lambda_b}^2- m_N^2) \gamma_\mu 
- (m_{\Lambda_b}+ m_N) \gamma_\mu \not\!q + (m_{\Lambda_b}+
m_N) q_\mu \Big] \nnb \\
\ar \Bigg[
{g_1^T(q^{2})\over m_{\Lambda_b}}
\Big[ \Big((m_{\Lambda_b}+ m_N)\gamma_\mu\gamma_5 + \gamma_\mu\not\!q
\gamma_5 + 2p_\mu \gamma_5\Big) q^2 \nnb \\
\ek \Big((m_{\Lambda_b}+ m_N) \not\!q\gamma_5  - (m_{\Lambda_b}^2- m_N^2 -
2 q^2)\gamma_5 \Big)q_\mu\Big] + g_2^T \Big[ -(m_{\Lambda_b}^2- m_N^2)
\gamma_\mu\gamma_5 \nnb \\
\ek (m_{\Lambda_b}- m_N)\gamma_\mu\not\!q\gamma_5
- 2 \not\!q\gamma_5 p_\mu + \Big((m_{\Lambda_b}- m_N) \gamma_5
+\not\!q\gamma_5 \Big) q_\mu\Big] \Bigg\}\nnb \\
\ar \frac{\lambda_{\Lambda_b^\ast}}{m_{\Lambda_b^\ast}^{2}-(p-q)^{2}}\Bigg\{
{\widetilde{f}_1^T(q^{2})\over m_{\Lambda_b^\ast}}
\Big[ \Big((m_{\Lambda_b^\ast}+ m_N)\gamma_\mu - \not\!q\gamma_\mu - 2p_\mu\Big) q^2 \nnb \\
\ek \Big((m_{\Lambda_b^\ast}+ m_N) \not\!q + (m_{\Lambda_b^\ast}^2- m_N^2 -
2 q^2)\Big)q_\mu\Big] + \widetilde{f}_2^T \Big[ (2 p_\mu -q_\mu)\not\!q  \nnb \\
\ar (m_{\Lambda_b^\ast}^2- m_N^2) \gamma_\mu 
- (m_{\Lambda_b^\ast}- m_N) \gamma_\mu \not\!q + (m_{\Lambda_b^\ast}-
m_N) q_\mu \Big] \nnb \\
\ar {\widetilde{g}_1^T(q^{2})\over m_{\Lambda_b^\ast}}
\Big[ \Big((m_{\Lambda_b^\ast} - m_N)\gamma_\mu\gamma_5 -
\gamma_\mu\not\!q\gamma_5 
- 2 \gamma_5 p_\mu\Big) q^2 \nnb \\
\ek \Big((m_{\Lambda_b^\ast} - m_N) \not\!q\gamma_5  + (m_{\Lambda_b^\ast}^2- m_N^2 -
2 q^2)\gamma_5 \Big)q_\mu\Big] + \widetilde{g}_2^T \Big[ (m_{\Lambda_b^\ast}^2- m_N^2)
\gamma_\mu\gamma_5 \nnb \\
\ek (m_{\Lambda_b^\ast}+ m_N)\gamma_\mu\not\!q\gamma_5
+ 2 \not\!q\gamma_5 p_\mu + (m_{\Lambda_b^\ast}+ m_N) \gamma_5 q_\mu
-\not\!q\gamma_5q_\mu \Big]\Bigg\}~.
%%%%%%%%%
\eea
We proceed now calculating the correlation functions (see Eq.
(\ref{ebrk03})) for the  $\Lambda_{b}(\Lambda_{b}^*)\rar N\ell^+\ell^- $
transitions from the QCD sides.
Note that, in the rest of the study the masses of the light
quarks are neglected. Moreover the external momenta $(p-q)$ and $q$ are
both space--like i.e., $(p-q)\ll m_b^2$ and $q\ll m_b^2$,
in order to justify the operator product expansion (OPE) near
the light cone $x^2\simeq 0$. The OPE is performed over the twist of the
nonlocal operators and it includes the nucleon distribution amplitudes (DAs).
Contracting the b--quark fields for the correlation functions, from the QCD
side we get,
\bea
\label{ebrk11}
\Pi^i_{\mu} \es \frac{i}{\sqrt{6}} \epsilon^{abc}\int d^4x e^{iqx}
\Big\{\Big[2 ( C )_{\alpha\gamma} (\gamma_5)_{\rho\tau}+( C
)_{\alpha\tau} (\gamma_5)_{\rho\gamma}+( C
)_{\tau\gamma} (\gamma_5)_{\alpha\rho}\Big] \nnb \\
\ar\beta\Big[2 (C \gamma_5
)_{\alpha\gamma}(I)_{\rho\tau}
+ (C \gamma_5 )_{\alpha\tau}(I)_{\rho\gamma}+(C \gamma_5
)_{\tau\gamma}(I)_{\alpha\rho} \Big]\Big\}  \nnb \\
\cp \Big(
\Gamma^i\Big)_{\sigma\beta}\Big(S_b(-x)\Big)_{\tau\sigma}
\langle 0 |  u_\alpha^a(0)
d_\beta^b(x) d_\gamma^c(0) | N (p)\rangle~,
\eea
where
\bea
\Gamma^i = \left\{
  \begin{array}{ll}
   \gamma_{\mu} (1-\gamma_5) & i=I~,\\
    i  \sigma_{\mu\nu} (1+\gamma_5)q^\nu & i=II~.
  \end{array}
\right. \nnb
\eea
The heavy quark operator $S_Q (x)$ is obtained in \cite{Rbrk16}, whose
expression is given as,
\bea
\label{ebrk12}
 S_Q (x)& =&  \frac{m_{Q}^{2}}{4\pi^{2}}\frac{K_{1}(m_{Q}\sqrt{-x^2})}
{\sqrt{-x^2}}-i \frac{m_{Q}^{2}\not\!x}{4\pi^{2}x^2}K_{2}
(m_{Q}\sqrt{-x^2})  \nnb \\
\ek i g_s \int \frac{d^4 k}{(2\pi)^4} e^{-ikx} \int_0^1 dv 
\Bigg[\frac{\not\!k + m_Q}{( m_Q^2-k^2)^2} G^{\mu\nu}(vx)
\sigma_{\mu\nu} + \frac{1}{m_Q^2-k^2} v x_\mu
G^{\mu\nu} \gamma_\nu \Bigg]~.
\eea
where $K_i$ are the modified Bessel functions, and $G_{\mu\nu}$
is the gluon field strength tensor.
It follows from Eq. (\ref{ebrk11}) that in order to calculate the correlator function
the matrix element
$\epsilon^{abc}\langle 0 |  u_\alpha^a(0)
d_\beta^b(x)   d_\gamma^c(0) |  N (p)\rangle~,\nnb$
of the three quark field operators between vacuum and nucleon
near light cone $x^2\rightarrow 0$ is needed.
This matrix element is parameterized in terms of the nucleon DAs
and is given in \cite{Rbrk06,Rbrk07} (for more details about the nucleon
DAs, see also \cite{Rbrk20}).
Substituting the parametrization of the matrix element
$ \epsilon^{abc}\langle 0 | u_\alpha^a(0)d_\beta^b(x) d_\gamma^c(0)| N (p)\rangle$
in terms of the nucleon DAs and the heavy $b$--quark propagator, the correlation
function from the QCD side can be calculated straightforwardly.

In order to suppress the higher states and continuum contributions 
we perform the Borel transformation over $-(p-q)^2$
to the expressions of correlation function from the hadronic and the QCD sides, and 
matching the coefficients of the relevant structures we get the following sum 
rules for the form factors:

(a) For $\gamma_\mu$ current
\bea
\label{ebrk13}
2\lambda_{\Lambda_b}f_{1}(q^2)e^{-m_{\Lambda_b}^2/M^{2}}-2\lambda_{\Lambda_b^\ast}
\widetilde{f}_{1}(q^2)
e^{-m_{\Lambda_b^\ast}^2/M^{2}}\es\Pi_{1}^{(I)B}(p,q)\nnb \\
-2\lambda_{\Lambda_b}\frac{f_{2}(q^2)}{m_{\Lambda_b}}e^{-m_{\Lambda_b}^2/M^{2}}+
2\lambda_{\Lambda_b^\ast}\frac{\widetilde{f}_2(q^2)}{m_{\Lambda_b^\ast}}
e^{-m_{\Lambda_b^\ast}^2/M^{2}}\es \Pi_{2}^{(I)B}(p,q) \nnb \\
\lambda_{\Lambda_b}e^{-m_{\Lambda_b}^2/M^{2}}\Big((m_{\Lambda_b}-m_N)(f_{1}(q^2)
-\frac{f_{2}(q^2)}{m_{\Lambda_b}}
(m_{\Lambda_b}+m_N))\Big)+\nnb \\
\lambda_{\Lambda_b^\ast}e^{-m_{\Lambda_b^\ast}^2/M^{2}}\Big((m_{\Lambda_b^\ast}+m_N)
(\widetilde{f}_1(q^2)+
\frac{\widetilde{f}_2(q^2)}{m_{\Lambda_b^\ast}}(m_{\Lambda_b^\ast}-m_N))\Big)\es
\Pi_{3}^{(I)B}(p,q)\nnb \\
\lambda_{\Lambda_b}e^{-m_{\Lambda_b}^2/M^{2}}\Big(f_{1}(q^2)-
\frac{f_{2}(q^2)}{m_{\Lambda_b}}(m_{\Lambda_b}+
m_N)\Big)+\nnb \\
\lambda_{\Lambda_b^\ast}e^{-m_{\Lambda_b^\ast}^2/M^{2}}\Big(-\widetilde{f}_1(q^2)-
\frac{\widetilde{f}_2(q^2)}{m_{\Lambda_b^\ast}}(m_{\Lambda_b^\ast}-m_N)\Big)\es
\Pi_{4}^{(I)B}(p,q)\nnb \\
\lambda_{\Lambda_b}e^{-m_{\Lambda_b}^2/M^{2}}\Big(-2f_{1}(q^2)+
\frac{(f_{2}(q^2)+ f_{3}(q^2))}{m_{\Lambda_b}}(m_{\Lambda_b^\ast}+m_N)\Big)+\nnb \\
\lambda_{\Lambda_b^\ast}e^{-m_{\Lambda_b^\ast}^2/M^{2}}\Big(2\widetilde{f}_1(q^2)+
\frac{(\widetilde{f}_2(q^2)+\widetilde{f}_3(q^2))}{m_{\Lambda_b^\ast}}
(m_{\Lambda_b^\ast}-m_N)\Big)\es\Pi_{5}^{(I)B}(p,q)\nnb \\
\frac{\lambda_{\Lambda_b}}{m_{\Lambda_b}}e^{-m_{\Lambda_b}^2/M^{2}}
\Big(f_{2}(q^2)-f_{3}(q^2)\Big)-
\frac{\lambda_{\Lambda_b^\ast}}{m_{\Lambda_b^\ast}}e^{-m_{\Lambda_b^\ast}^2/M^{2}}
\Big(\widetilde{f}_2(q^2)-
\widetilde{f}_3(q^2)\Big)\es\Pi_{6}^{(I)B}(p,q),
\eea
where, superscript $I$ represents transition current $\gamma_\mu$. Here,
$\Pi_{1}^{(I)B}(p,q)$, $\Pi_{2}^{(I)B}(p,q)$, $\Pi_{3}^{(I)B}(p,q)$, 
$\Pi_{4}^{(I)B}(p,q)$, $\Pi_{5}^{(I)B}(p,q)$, and $\Pi_{6}^{(I)B}(p,q)$
are the invariant functions for the Lorentz structures, $p_{\mu}$, 
$p_{\mu}\not\!q$, $\gamma_{\mu}$,$\gamma_{\mu}\not\!q$, $q_\mu$, and 
$q_{\mu}\not\!q$ structures, respectively.

The results for $\gamma_\mu\gamma_5$ current are obtained from Eq.
(\ref{ebrk13}) with the following replacements:
$f_i\rightarrow g_i$, $\widetilde{f}_i\rightarrow \widetilde{g}_i$, 
$m_N\rightarrow -m_N$, and $\Pi_i^{(I)B}\rightarrow \Pi_i^{(I)1B}$.

The sum rules for the form factors induced by the $i\sigma_{\mu\nu}q^\nu$ 
current we get:

\bea
\label{ebrk14}
-2\lambda_{\Lambda_b}f_{2}^T(q^2)e^{-m_{\Lambda_b}^2/M^{2}}+2\lambda_{\Lambda_b^\ast}
\widetilde{f}_{2}^T(q^2)e^{-m_{\Lambda_b^\ast}^2/M^{2}}\es\Pi_{1}^{(II)B}(p,q)\nnb \\
\lambda_{\Lambda_b}e^{-m_{\Lambda_b}^2/M^{2}}\Big(-\frac{f_{1}^T(q^2)}{m_{\Lambda_b}}(m_{\Lambda_b}-m_N)+
f_{2}^T(q^2)\Big)+\nnb \\
\lambda_{\Lambda_b^\ast}e^{-m_{\Lambda_b^\ast}^2/M^{2}}\Big(-\frac{\widetilde{f}_1^T(q^2)}{m_{\Lambda_b^\ast}}
(m_{\Lambda_b^\ast}+m_N)-\widetilde{f}_2^T(q^2)\Big)\es\Pi_{2}^{(II)B}(p,q)\nnb \\
\lambda_{\Lambda_b}e^{-m_{\Lambda_b}^2/M^{2}}\Big(\frac{f_{1}^T(q^2)}{m_{\Lambda_b}}
(m_{\Lambda_b}^2-m_N^2-2q^2)+f_{2}^T(q^2)(m_{\Lambda_b}+m_N)\Big)+\nnb \\
\lambda_{\Lambda_b^\ast}e^{-m_{\Lambda_b^\ast}^2/M^{2}}\Big(-\frac{\widetilde{f}_1^T(q^2)}{m_{\Lambda_b^\ast}}
(m_{\Lambda_b^\ast}^2-m_N^2-2q^2)+\widetilde{f}_2^T(q^2)(m_{\Lambda_b^\ast}-m_N)\Big)\es\Pi_{3}^{(II)B}(p,q)\nnb \\
\lambda_{\Lambda_b}e^{-m_{\Lambda_b}^2/M^{2}}(m_{\Lambda_b}-m_N)\Big(\frac{f_{1}^T(q^2)}{m_{\Lambda_b}}q^2-
f_{2}^T(q^2)(m_{\Lambda_b}+m_N)\Big)+\nnb \\
\lambda_{\Lambda_{b}^\ast}e^{\frac{-m_{\Lambda_{b}^\ast}^2}{M^{2}}}(m_{\Lambda_{b}^\ast}+m_N)
\Big(\frac{\widetilde{f}_1^T(q^2)}{m_{\Lambda_b^\ast}}q^2+\widetilde{f}_2^T(q^2)(m_{\Lambda_b^\ast}-m_N)
\Big)\es\Pi_{4}^{(II)B}(p,q)~,
\eea
where $\Pi_{1}^{(II)B}$, $\Pi_{2}^{(II)B}$, $\Pi_{3}^{(II)B}$, and $\Pi_{4}^{(II)B}$
are the invariant functions for the structures $p_{\mu}\not\!q$,
$q_{\mu}\not\!q$, $q_\mu$, and $\gamma_{\mu}$, respectively.
The sum rules for $i\sigma_{\mu\nu}q^\nu\gamma_5 $ transition current can be 
obtained from equation (14) by making the replacements $f_i^T\rightarrow 
g_i^T$, $\widetilde{f}_i^T\rightarrow \widetilde{g}_i^T$, $m_N\rightarrow -m_N$, 
and $\Pi_i^{(II)B}\rightarrow \Pi_i^{(II)1B}$. The explicit form of these
invariant functions are quite lengthy, and for this reason we do not present
them in this work.

Few remarks about calculations of the Borel transformation from the $QCD$ side are in
order. After performing the Fourier transformation the invariant amplitudes get the 
following form
\bea
\label{ebrk15}
\Pi_{i}\Big((p-q)^2,q^2\Big)\es\int_{0}^{1}dx\frac{D((p-q)^2,q^2)}{\Delta^n}~,
\eea
where the denominator is given by  $\Delta=m_b^2-(-xp+q)^2$=$m_b^2-\bar{x}q^2+
x\bar{x}m_N^2-x(p-q)^2$, and $\bar{x}=1-x$.
In order to perform Borel transformation we rewrite denominator in the form
\bea
\label{ebrk16}
\Delta=x\Big(s(x)-(p-q)^2\Big)~,
\eea
where $s(x)=(m_b^2-\bar{x}q^2+x\bar{x}m_N^2)/x$.
Following this replacement, the Borel transformation and continuum subtraction 
are performed \cite{Rbrk06}:
\bea
\label{ebrk17}
\int dx\frac{D(x)}{\Delta} &\rightarrow& \int_{x_{0}}^{1}\frac{dx}{x} D(x)
e^{\frac{-s(x)}{M^{2}}}\nnb \\
\int dx\frac{D(x)}{\Delta^2} &\rightarrow& \frac{1}{M^2} \int_{x_{0}}^{1}
\frac{dx}{x^2}D(x)e^{\frac{-s(x)}{M^{2}}}+ \frac{D(x_{0})
e^{\frac{-s_{0}}{M^{2}}} }{m_{b}^2+x_{0}^{2}m_{N}^2-q^2}\nnb \\
\int dx\frac{D(x)}{\Delta^3} &\rightarrow& \frac{1}{2M^4} \int_{x_{0}}^{1}
\frac{dx}{x^3}D(x)e^{\frac{-s(x)}{M^{2}}}+\frac{1}{2M^2} \frac{D(x_{0})
e^{\frac{-s_{0}}{M^{2}}} }{x_{0}(m_{b}^2+x_{0}^{2}m_{N}^2-q^2)}\nnb \\
\ek\frac{1}{2} \frac{x_{0}^2e^{\frac{-s_{0}}{M^{2}}}}{m_{b}^2+x_{0}^{2}
m_{N}^2-q^2}\frac{d}{dx}\Big( \frac{D(x)}{x(m_{b}^2+x^{2}m_{N}^2-q^2)}
\Big)\Big|_{x=x_{0}}~,
\eea
where $x_{0}$ is the solution of the equation
\bea
\label{ebrk18}
s_{0}=\frac{m_b^2-\bar{x}q^2+x\bar{x}m_N^2}{x}.
\eea

Solving Eqs. (\ref{ebrk13}) and (\ref{ebrk14}) we obtain the desired sum rules for
the transition form factors $f_{i}$, $g_{i}$, $f_{i}^T$, $g_{i}^T$,
$\widetilde{f}_i$, $\widetilde{g}_i$, $\widetilde{f}_i^T$, and $\widetilde{g}_i^T$.

One can easily see that, the expressions of the form factors contain the
residues $\lambda_{\Lambda_b}$ and $\lambda_{\Lambda_b^\ast}$ of the $\Lambda_b$
and $\Lambda_b^\ast$ baryons, respectively. These residues are determined from
the analysis of the following two--point correlation function,
\bea
\label{ebrk19}
T=i\int d^{4}xe^{iqx}\langle 0 \mid T\{ \eta_{\Lambda_{b}}(x)\bar{\eta}_{\Lambda_{b}}
\}(0) \mid  0\rangle=T_{1}(q^2)\not\!q+T_{2}(q^2)~.
\eea
Note that this correlator is used to calculate the residue of the $\Lambda_b$
baryon when for the pseudoscalar-- and axial--vector currents are used \cite{Rbrk11}.
We also use this correlator to calculate the residues of the
$\Lambda_b(\Lambda_b^\ast)$ baryons by using the most general form of the 
interpolating current of the $\Lambda_b$ baryon. Following the standard
method, i.e., performing the Borel transformation and continuum
subtraction procedures, we obtain, 
\bea
\label{ebrk20}
\lambda_{\Lambda_b}e^{-m_{\Lambda_b}^2/M^{2}}+\lambda_{\Lambda_b^\ast}
e^{-m_{\Lambda_b^\ast}^2/M^{2}}\es T_{1}^{B}\nnb \\
\lambda_{\Lambda_b}m_{\Lambda_b}e^{-m_{\Lambda_b}^2/M^{2}}-
\lambda_{\Lambda_b^\ast}m_{\Lambda_b^\ast}e^{-m_{\Lambda_b^\ast}^2/M^{2}}\es T_{2}^{B},
\eea
where, $\lambda_{\Lambda_b}(\lambda_{\Lambda_b^\ast})$, and $m_{\Lambda_b}(m_{\Lambda_b^\ast})$
are the residues and masses of $\Lambda_{b}(\widetilde{\Lambda}_{b})$ baryons.
Solving these two equations for the residues $\lambda_{\Lambda_b}$ and $\lambda_{\Lambda_b^\ast}$
we obtain,
\bea
\label{ebrk21}
\lambda_{\Lambda_b} \es
{e^{m_{\Lambda_b}^2/M^{2}}\over m_{\Lambda_b}+m_{\lambda_b^\ast}}
\Big(m_{\Lambda_b^\ast}T_{1}^{B}+T_{2}^{B}\Big),\\
\label{ebrk22}
\lambda_{\Lambda_b^\ast}\es {e^{m_{\Lambda_b^\ast}^2/M^{2}}\over m_{\Lambda_b}+m_{\lambda_b^\ast}} 
\Big(m_{\Lambda_b}T_{1}^{B}-T_{2}^{B}\Big).
\eea
The expressions of the invariant functions $T_{1}^{B}$, and $T_{2}^{B}$ can be
obtained from the results presented in  \cite{Rbrk17}.

\section{{Numerical Analysis}}

In this section we present the numerical results of the form factors of the
rare $\Lambda_{b}(\Lambda_{b}^\ast)\rar N\ell^+\ell^- $ decays, and their total
decay rates and branching ratios. The main input parameters used in the
numerical calculations are,
$m_N= 0.938~GeV$, $m_{\Lambda_{b}} = 5.620~ GeV$, and $m_{\Lambda_{b}^\ast} =
5.920~ GeV$ \cite{Rbrk18}. For the mass of the $b$--quark, we take its MS mass 
value $\bar{m}_{b} = (4.16\pm 0.03)~GeV$ \cite{Rbrk18},
$\uu(1~GeV) = \dd(1~GeV) = -(246^{+28}_{-19} MeV)^3$.

The input parameters entering the DAs of nucleon are taken from
\cite{Rbrk05,Rbrk06}, whose values are,
\bea
\label{ebrk23}
f_N\es(5.0\pm0.5)\times10^{-3}~ \mbox{GeV}^2~,\hspace{2.5cm}
\lambda_1=-(27\pm 9)\times10^{-3}~ \mbox{GeV}^2~,\nnb\\
\lambda_2\es(54\pm 19)\times10^{-3}~ \mbox{GeV}^2~,\hspace{2.6cm}A_1^u=0.13,\nnb\\
V_2^d\es 0.30~,\hspace{6.0cm}f_1^d=0.33,\nnb\\
f_1^u\es 0.09~,\hspace{6.0cm}f_2^d=0.25.
\eea

As has already been noted in further numerical analysis the value of
residues $\lambda_{\Lambda_b} (\lambda_{\Lambda_b^\ast})$ of the 
$\Lambda_b(\Lambda_b^\ast$ baryons are needed.
In this regard, the mass sum rules for the $\Lambda_b(\Lambda_b^\ast$
baryons for the most general form of the interpolating current contain 
three auxiliary parameters, namely, Borel mass parameter $M^2$,
continuum threshold $s_0$, and the arbitrary parameter $\beta$.
The working region of $M^2$ for the residue is determined by using the standard
criteria, i.e., the power corrections and continuum contributions should be 
suppressed at the chosen values of $s_0$ and $\beta$. As the result of these 
requirements the working region of the Borel mass parameter is found to be
$4 ~GeV^{2}\leq M^{2}\leq 6~ GeV^{2}$.
The continuum threshold can be obtained from the condition that the mass sum rules
reproduce the lowest baryon mass with an accuracy of $10\%$, for a given value
of $\beta$. The numerical analysis performed in this regard, determines the
working region of the continuum threshold to be $s_0=(40\pm 1)GeV^2$.
Finally, in order to find the working region for the parameter $\beta$,
we studied the dependence of $\Lambda_b (\Lambda_b^\ast)$ on $ \cos\theta$,
where $tan\theta=\beta$. We observe that the residue demonstrates a good
stability to the variation of $ \cos\theta$ in the domain  
$-1\leq \cos\theta\leq -0.5$, within an uncertainty of less than $4\%$.
Therefore, in the proceeding analysis we shall use $\beta=-1$.
Taking into account of the working regions of the aforementioned parameters
$M^2$, $s_0$ and $\beta$, the values of the residues $\lambda_{\Lambda_b}$ and
$\lambda_{\Lambda_b^\ast}$ which we shall use our analysis are found to be,
\bea
\label{ebrk24}
\lambda_{\Lambda_b}\es(6.5\pm 1.5)\times10^{-2}~ \mbox{GeV}^3~,\nnb\\
\lambda_{\Lambda_b^\ast}\es(7.5\pm 2.0)\times10^{-2}~ \mbox{GeV}^3~.
\eea

We now turn our attention to the calculation of the
$\Lambda_{b}(\Lambda_b^\ast)\to N$
transition form factors. The working region of $M^2$ for these form factors
is determined in accordance with the aforementioned requirement, i.e sufficient
suppression of the power correction and continuum contributions. Our analysis 
shows that these conditions are simultaneously satisfied when $M^2$ lies in 
the domain $15~GeV^{2}\leq M^{2}\leq 25~ GeV^{2}$. For the continuum threshold 
and the arbitrary parameter $\beta$, we use $s_0=(40\pm 1)GeV^2$ and 
$\beta=-1$, respectively.

The LCSR predictions, unfortunately, do not work in the entire physical region.
The prediction of LCSR for the form factors are reliable up to $q^{2}=(10-11)~ GeV^{2}$.
Since at large $q^2$ the contributions of higher twists become sizable and 
convergence of OPE is questionable.
In order to extend the LCSR prediction for the form factors to the entire physical domain
$q^{2}=(m_{\Lambda_{b}}-m_N)^2~ GeV^{2}$ we use the z-series parametrization 
suggested in \cite{Rbrk19}, where
\begin{equation}
\label{Rbrk25}
z(q^2, t_{0}) = \frac{\sqrt{t_{+}-q^2}-\sqrt{t_{+}-t_0}}{\sqrt{t_+-q^2}+
\sqrt{t_+-t_0}}~,
\end{equation}
with $t_0 = q^2_{\rm max} = (m_{\Lambda_b} - m_N)^2$, $t_+ = (m_B + m_\pi)^2$.

The best parametrization of the form factors, with which the predictions of
the LCSR are reproduced with a high accuracy in the $q^{2}\leq 11~ GeV^{2}$
region, is given as
\begin{equation}
\label{Rbrk26}
f(q^2) = \frac{1}{1-q^2/(m_{\rm pole}^f)^2} \big[ a_0^f + a_1^f\:z(q^2,t_0) +
a_2^f\:[z(q^2,t_0)]^2 \big]~.
\end{equation}
For the pole masses we use,
\bea
\label{Rbrk27}
 m_{pole} = \left\{
  \begin{array}{rl}
  m_{B^\ast} = 5.325~ GeV &  \!\mbox{for the form factors~~} f_1, f_2, f_{1}^T~,
f_{2}^T;\widetilde{g}_1, \widetilde{g}_2, \widetilde{g}_{1}^T~,
\widetilde{g}_{2}^T \\
    m_{B_1} = 5.723~ GeV & \!\mbox{for the form factors~~} g_1, g_2, g_{1}^T~, 
g_{2}^T; \widetilde{f}_1, \widetilde{f}_2, \widetilde{f}_{1}^T~, 
\widetilde{f}_{2}^T \\
     m_{B_0} = 5.749~ GeV & \!\mbox{for the form factors~~} f_3~; 
\widetilde{g}_3\\
    m_{B} = 5.280~ GeV  & \!\mbox{for the form factors  } g_3~;
\widetilde{f}_3 \\
  \end{array}
\right.
\eea

Our analysis predicts the following  values of fit parameters 
$a^f_{0}$, $a^f_1$ and $a^f_2$ for the 
$\Lambda_b \to N \ell^+ \ell^-$ and $\Lambda_b^\ast \to N \ell^+ \ell^-$
form factors, respectively, which are presented in  Tables 1 and 2

% .........................................................

\begin{table}[h]

\renewcommand{\arraystretch}{1.3}
\addtolength{\arraycolsep}{-0.5pt}
\small
$$
\begin{array}{|l|c|c|c|c|}
\hline \hline
        &        f_i(0)    &         a_0       &        a_1      &        a_2       \\  \hline
f_1     & -0.075 \pm 0.005 &    0.17 \pm 0.03  &  -1.56 \pm 0.09 &   2.46 \pm 1.00  \\
f_2     &  0.11  \pm 0.01  &    0.79 \pm 0.05  &  -3.15 \pm 1.30 &   3.35 \pm 1.40  \\
f_3     &  0.08  \pm 0.00  &    0.78 \pm 0.06  &  -3.5  \pm 1.3  &   4.3  \pm 1.4   \\
g_1     & -0.090 \pm 0.002 &    0.21 \pm 0.04  &  -1.96 \pm 0.90 &   3.18 \pm 1.30  \\
g_2     &  0.08  \pm 0.003 &    0.80 \pm 0.06  &  -3.5  \pm 1.3  &   4.1  \pm 1.3   \\
g_3     &  0.14  \pm 0.04  &    1.10 \pm 0.03  &  -4.6  \pm 1.4  &   5.0  \pm 1.6   \\
f_1^T   &  0.11  \pm 0.02  &   11    \pm 3     & -66    \pm 15   & 100    \pm 25    \\
f_2^T   & -0.13  \pm 0.03  &  - 2.6  \pm 1.0   &  14    \pm 3    &  19    \pm 6     \\
g_1^T   &  0.12  \pm 0.03  &    2.5  \pm 0.9   & -13    \pm 3    &  17    \pm 6     \\
g_2^T   & -1.8   \pm 0.03  &  - 1.0  \pm 0.2   & -0.60  \pm 0.14 &   0.10 \pm 0.02  \\
\hline \hline
\end{array}
$$
\caption{Form factors of the $\Lambda_b \to \ell^+ \ell^-$ decay}
\renewcommand{\arraystretch}{1}
\addtolength{\arraycolsep}{-1.0pt}
\end{table}

% .........................................................

\begin{table}[h]

\renewcommand{\arraystretch}{1.3}
\addtolength{\arraycolsep}{-0.5pt}
\small
$$
\begin{array}{|l|c|c|c|c|}
\hline \hline
                    & \widetilde{f}_i(0)   &         a_0           &        a_1       &        a_2        \\  \hline
\widetilde{f}_1     &  -0.002 \pm 0.001    &    0.60   \pm 0.20    &  -2.71 \pm 0.5        &  3.04  \pm 0.70   \\
\widetilde{f}_2     &  -0.040 \pm 0.001    &   -0.36   \pm 0.10    &   1.19 \pm 0.24  &  -1.0   \pm 0.2   \\
\widetilde{f}_3     &  -0.052 \pm 0.00 2   &    0.0085 \pm 0.0002  &  -0.53 \pm 0.12  &   1.1   \pm 0.23  \\
\widetilde{g}_1     &  -0.030 \pm 0.006    &    -0.11  \pm 0.02    &   0.28 \pm 0.06  &  -0.230 \pm 0.045 \\
\widetilde{g}_2     &  -0.044 \pm 0.002    &    -0.11  \pm 0.02    &   0.11 \pm  0.02 &   0.19  \pm 0.03  \\
\widetilde{g}_3     &  -0.020 \pm - 0.004  &    -0.085 \pm 0.002   & 0.24   \pm 0.05  &   0.170 \pm 0.032 \\
\widetilde{f}_1^T   &  0.070 \pm 0.003     &   -11     \pm         &  52    \pm 1     &  -65    \pm 15    \\
\widetilde{f}_2^T   &  -0.030 \pm 0.004    &   -2.7    \pm 0.6     &  12.5  \pm 2.5   &  -15    \pm 3     \\
\widetilde{g}_1^T   &  -0.080 \pm 0.002    &   -3.0    \pm 0.7     & -13.5  \pm 3.5   &  -15.3  \pm 3.5   \\
\widetilde{g}_2^T   &  -0.060 \pm 0.002    &   0.040 \pm 0.008     & -0.90  \pm 0.05  &  -1.8   \pm 0.4   \\
\hline \hline
\end{array}
$$
\caption{The same as Table 1, but for the $\Lambda_b^\ast \to \ell^+ \ell^-$ decay}
\renewcommand{\arraystretch}{1}
\addtolength{\arraycolsep}{-1.0pt}
\end{table}

Having the transition form factors at hand, we can now calculate the branching ratio of 
the $\Lambda_{b}(\Lambda_b^\ast)\to N\ell^{+}\ell^{-}$ decays. The effective Hamiltonian for
the $b\rightarrow  d\ell ^{+}\ell ^{-}$ transition has the following form
\cite{Rbrk20}:
\begin{equation}
\label{Rbrk28}
H_{eff}=\frac{4G_{F}}{\sqrt{2}}V_{tb}V^{* }_{td} \left \{\sum\limits_{i=1}^{10}
C_{i}(\mu)O_{i}(\mu)+  V_{ub}V^{* }_{ud}\sum\limits_{i=1}^{2}C_{i}(O_{i}-O_{i}^u)
\right\}~,
\end{equation}
where, $C_{i}$ are the Wilson coefficients and set $O_i$ represent local operators.
The matrix element responsible for the $\Lambda_{b}\to N\ell^{+}\ell^{-}$ transition can
be obtained from the effective Hamilton by sandwiching it between $\Lambda_{b}$ and $ N$
states, after which the matrix element responsible for $\Lambda_{b}\to N\ell^{+}\ell^{-}$ 
decay takes the following form,
\bea
\label{Rbrk29}
{\cal M}\es\frac{G_F\alpha}{\sqrt{2}\pi}|V_{td}^\ast V_{tb}|\Bigg\{ \bar{u}_{N}(p)
\Bigg[\gamma_\mu F_{1}+\frac{i\sigma_{\mu\nu}q^\nu}{m_{\Lambda_{b}}}F_{2}- 
\frac{q_\mu}{m_{\Lambda_{b}}}F_{3}-\gamma_\mu \gamma_5 G_{1}
+ \frac{i\sigma_{\mu\nu}q^\nu}{m_{\Lambda_{b}}}\gamma_5 G_{2} \nnb \\
\ek \frac{q_\mu}{m_{\Lambda_{b}}}\gamma_5 G_{3}\Bigg]u_{\Lambda_b}(p+q)
(\bar \ell\gamma^\mu \ell) + \bar{u}_{N}(p)\Bigg[\gamma_\mu F_4+
\frac{i\sigma_{\mu\nu}q^\nu}{m_{\Lambda_{b}}} F_5- 
\frac{q_\mu}{m_{\Lambda_{b}}} F_6
-\gamma_\mu \gamma_5 G_4 \nnb \\
\ar\frac{i\sigma_{\mu\nu}q^\nu}{m_{\Lambda_{b}}}
\gamma_5 G_5- \frac{q_\mu}{m_{\Lambda_{b}}}\gamma_5 G_6 \Bigg]
u_{\Lambda_b}(p+q)(\bar \ell\gamma^\mu \gamma_5 \ell)\Bigg\},
\eea
where,
\bea
\label{Rbrk30}
F_1 \es c_9f_1-\frac{2m_b }{m_{\Lambda_{b}}}c_7f_{1}^T~,\nnb\\
F_2 \es c_9f_2+\frac{2m_b }{q^2}m_{\Lambda_{b}}f_{2}^T~,\nnb\\
F_3 \es c_9f_3-\frac{2m_b }{q^2}c_7(m_{\Lambda_{b}}-m_N)f_{1}^T~,\nnb\\
G_1 \es c_9g_1-\frac{2m_b }{m_{\Lambda_{b}}}c_7g_{1}^T~,\nnb\\
G_2 \es c_9g_2+\frac{2m_b }{q^2}m_{\Lambda_{b}}g_{2}^T~,\nnb\\
G_3 \es c_9g_3-\frac{2m_b }{q^2}c_7(m_{\Lambda_{b}}+m_N)g_{1}^T~,\nnb\\
F_4 \es c_{10}f_1~,\nnb\\
F_5 \es c_{10}f_2~,\nnb\\
F_6 \es c_{10}f_3~,\nnb\\
G_4 \es c_{10}g_1~,\nnb\\
G_5 \es c_{10}g_2~,\nnb\\
G_6 \es c_{10}g_3~.
\eea

The matrix element for the $\Lambda_b^\ast \rar N \ell^+\ell^-$
transition cam be obtained from the matrix element for the
$\Lambda_b \rar N \ell^+\ell^-$ by first making the following replacements:
$F_i \rar \widetilde{G}_i$, $G_i \rar \widetilde{F}_i$, $ m_N \rar - m_N$,
$m_{\Lambda_b} \rar m_{\Lambda_b^\ast}$; and further making the
following ones:
$f_i \rar \widetilde{f}_i$, $f_i^T \rar \widetilde{f}_i^T$,
$g_i \rar \widetilde{g}_i$, $g_i^T \rar \widetilde{g}_i^T$.

Our final goal is to calculate the differential width of 
the $\Lambda_{b}^{*}\to \Lambda\ell^{+}\ell^{-}$ decay, whose expression is
given as
\bea
\label{Rbrk33}
\frac{d\Gamma(s)}{ds} = \frac{G^2\alpha^2_{em} m_{\Lambda_b}}{4096 \pi^5}|
V_{tb}V_{td}^*|^2 v \sqrt{\lambda(1,r,s)} \, \Bigg[  \Gamma_1(s) +
\frac{1}{3} \Gamma_2(s)\Bigg]~,
\eea
where $\alpha$ is the fine structural constant, $v_\ell=\sqrt{1-
4 m_\ell^2/q^2}$ is the lepton velocity,
$\lambda(1, r,  s)=1+r^2+s^2-2r-2 s-2r s$,
$s=q^2/m_{\Lambda_{b}}$, and  $r=m_{\Lambda}^2/m_{\Lambda_{b}}^2$,
and the expressions of $\Gamma_1(s)$ and $\Gamma_2(s)$ are given in the Appendix.

We also calculate the differential decay width for the
$\Lambda_{b}^{*}\to N\ell^{+}\ell^{-}$ transition whose expression 
can easily be obtained from the differential decay width of the 
$\Lambda_{b}\to N\ell^{+}\ell^{-}$ transition with the help of
appropriate replacements.

The differential branching ratios for the $\Lambda_{b}\to N\mu^{+}\mu^{-}$ ,
$\Lambda_{b}\to N\tau^{+}\tau^{-}$ , $\Lambda_{b}^\ast\to N\mu^{+}\mu^{-}$, 
and $\Lambda_{b}^\ast\to N\tau^{+}\tau^{-}$ decays at $s_{0}=40~ GeV^2$ and 
$M^2=25~GeV^2$ are presented in figures 1, 2, 3, and 4 respectively.

In order to calculate the branching ratios of the
$\Lambda_{b}(\Lambda_{b}^{*})\to N\ell^{+}\ell^{-}$ transitions,
the differential decay width of the respective decays should be
integrated over $s$ in the domain $4m_{\ell}^2/m_{\Lambda_b}^2\leq 
s\leq  (1-\sqrt{r})^2$, and this results 
should be multiplied with the the total life time of the
$\Lambda_{b}(\Lambda_{b}^{*})$ baryons, reactively. 
In these calculations we neglect the long distance contributions coming from
the  $J/\Psi$ family to the $c_{9}$. The values of Wilson coefficients are taken 
from \cite{Rbrk20}: $c_{9}=4.34$, $c_{10}=-4.669$, and $c_{7}=-0.313$.
As the result of these considerations, 
the branching ratios for the $\Lambda_{b}\to N \ell^+ \ell^-~(e,~\mu,~\tau)$
transitions are calculated to have the values $Br(\Lambda_{b}
\to Ne^{+}e^{-})=(8\pm2).10^{-8}$ ,  $Br(\Lambda_{b}\to N\mu^{+}\mu^{-})=
(7\pm2).10^{-8}$ , and $Br(\Lambda_{b} \to N\tau^{+}\tau^{-})=(2\pm
0.4).10^{-8}$, respectively.

In determining the branching ratios of the $\Lambda_{b}^\ast\to
N\ell^{+}\ell^{-}~(e,~\mu,~\tau)$ decay, the life time of 
of the $\Lambda_{b}^\ast$ is needed, which approximately has the same value
as that of the $\Lambda_{b}$ baryon. 
So multiplying the branching ratios of the $\Lambda_{b}\to N \ell^+
\ell^-~(e,~\mu,~\tau)$ with the factor $\tau(\Lambda_{b})/\tau(\Lambda_{b}^\ast)$
will yield more precise values for these transitions.

At the end of this section we would like to make few comments on the results of 
the $\Lambda_{b}\to N$ transition form factors with the ones existing 
in the literature.
The form factors of $\Lambda_{b}\to N\ell\nu$ transition has already been
calculated in framework of the LCSR in \cite{Rbrk20}, by taking 
the contributions of the $\Lambda_{b}$ and $\Lambda_{b}^\ast$ decays into
account. Our results for the form factors $f_{1}$, $f_{2}$, $g_{1}$, 
and $g_{2}$ are different compared to the ones presented in \cite{Rbrk20}.
This can be attributed to the fact that, in the present work, we have used different
form of the interpolating current than that used in \cite{Rbrk20}. 
As we have already noted, the results for the $\Lambda_{b}^\ast$ baryon are
very sensitive to the choice of the interpolating current. We have also checked that
if interpolating current presented in \cite{Rbrk20} were used, our results
on the form factors coincide with the predictions of the work \cite{Rbrk20}.
The $\Lambda_{b}\to N$ transition is studied in \cite{Rbrk21} in framework
of the LCSR, but without taking into the contributions 
of $\Lambda_{b}^\ast$ baryons into account. And also
the continuum subtraction procedure is performed rather in an inconsistent
manner. For this reason, our predictions for the branching ratios are a bit off
compared to the ones presented in \cite{Rbrk21}, due to the considerable
differences on the predictions of the form factors.
Moreover, the $\Lambda_b \to N\ell^{+}\ell^{-}$ decay is studied within
the relativistic quark--diquark picture in \cite{Rbrk22}.
When compared, our predictions on the relevant form factors are different
than those predicted in \cite{Rbrk22}, where the results for the branching ratios
are, approximately, two times smaller compared to our predictions.

The sum rules for the form factors can further be improved by taking into
account the $\alpha_s$ corrections to the DAs, and improvements on the 
input parameters present in these DAs. The results we obtain for
branching ratios of the the CKM suppressed $\Lambda_b \to N \ell^+\ell$ 
and $\Lambda_b^\ast \to N \ell^+\ell$ decays governed by the $b \to d \ell^+
\ell^-$ transition, give confidence that these decays can
be discovered at LHCb at near future.

\section{Conclusion}

In present work, we calculate the transition form factors of $\Lambda_b \to N 
\ell^+\ell^-$ decay in the framework of the LCSR. We take into account the 
contribution of negative parity $\Lambda_b^\ast$ baryon to the sum rules. 
Using the obtained results for the form factors we estimate the branching 
ratios of $\Lambda_b \to N \ell^+\ell^-$, and $\Lambda_b^\ast\to N \ell^+\ell^-$ decays.
We also compare our predictions on the form factors and branching ratios 
with the ones existing in the literature. From these results
we can conclude that these decays can be observed at the near 
future at LHCb.
\\
\\
\section*{Acknowledgments}
 One of the authors (T. Barakat) extends his
appreciation to the International Scientific Partnership Program 
ISPP at King Saud University for funding this research work through ISPP
No. 0038.

\newpage

\newpage

% ...........................................................

\section*{Appendix : Differential decay widths for the
$\Lambda_{b}\to N\ell^{+}\ell^{-}$ and $\Lambda_{b}^\ast\to
N\ell^{+}\ell^{-}$ transitions}
%\section*{}
\setcounter{equation}{0}
\setcounter{section}{0}

% ...........................................................

In this Appendix we present the differential decay widths for the
$\Lambda_{b}\to N\ell^{+}\ell^{-}$ and $\Lambda_{b}^\ast\to
N\ell^{+}\ell^{-}$ transitions. After lengthy, but straightforward
calculations, for the differential decay rate of the $\Lambda_{b}\to
N\ell^{+}\ell^{-}$ we get,

\bea
\frac{d\Gamma(s)}{ds} = \frac{G^2\alpha^2_{em} m_{\Lambda_b}}{4096 \pi^5}|
V_{tb}V_{td}^*|^2 v \sqrt{\lambda(1,r,s)} \, \Bigg[  \Gamma_1(s) +
\frac{1}{3} \Gamma_2(s)\Bigg]~, \nnb
\eea
where $s= q^2/m^2_{\Lambda_b}$, $r= m_{N}^2/m^2_{\Lambda_b}$,
$G_F = 1.17 \times 10^{-5}$ GeV$^{-2}$ is the Fermi
coupling constant, $v=\sqrt{1-4 m_\ell^2/q^2}$ is the
lepton velocity, and $\lambda(a,b,c)=a^2+b^2+c^2-2ab-2ac-2bc$ is the usual
triangle function. For the element of the CKM matrix
$\vel V_{tb}V_{td}^\ast \ver = (8.2 \pm 0.6)\times 10^{-3}$ has been used
\cite{Rbrk19}. The functions $\Gamma_1(s)$ and $\Gamma_2(s)$ are given as:

\bea
\Gamma_1(s) \es
%-----------------
% term 1
%
8  m_{\Lambda_b}^2 \Bigg\{
(1 - 2 \sqrt{r} + r - s) \left[4 m_\ell^2  + m_{\Lambda_b}^2  (1 + 2 \sqrt{r} + r + s)\right]
\ve F_1 \ve^2 \nnb \\
%---------------------
% term 2
\ek \Big[4 m_\ell^2  (1 - 6 \sqrt{r} + r - s) -  
        m_{\Lambda_b}^2  \Big( (1 - r)^2  - 4 \sqrt{r} s - s^2 \Big)\Big]
 \ve F_4 \ve^2 \nnb \\
%---------------------
% term 3
\ar (1 - 2 \sqrt{r} + r - s) 
    \Big[4 m_\ell^2  (1 + \sqrt{r})^2  + m_{\Lambda_b}^2  s (1 + 2 \sqrt{r} + r +
     s)\Big] \ve F_2 \ve^2 \nnb \\
%---------------------
% term 4
\ar m_{\Lambda_b}^2  s \Big[(-1 + r)^2  - 4 \sqrt{r} s - s^2 \Big] v^2
  \ve F_4 \ve^2 \nnb \\
%---------------------
% term 5
\ar 4 m_\ell^2  (1 + 2 \sqrt{r} + r - s) s \ve F_6 \ve^2 \nnb \\
%---------------------
% term 6                                        
\ar (1 + 2 \sqrt{r} + r - s) \Big[4 m_\ell^2  + m_{\Lambda_b}^2  (1 - 2 \sqrt{r} + r + s) \Big] 
    \ve G_1 \ve^2 \nnb \\
%---------------------
% term 7
\ek \Big[ 4 m_\ell^2  (1 + 6 \sqrt{r} + r - s) - 
        m_{\Lambda_b}^2  \Big( (1 - r)^2  + 4 \sqrt{r} s - s^2 \Big)\Big]
 \ve G_4 \ve^2 \nnb \\
%---------------------
% term 8
\ar (1 + 2 \sqrt{r} + r - s) 
    \Big[4 m_\ell^2  (1 - \sqrt{r})^2  + m_{\Lambda_b}^2  s (1 - 2 \sqrt{r} + r +
s)\Big] \ve G_2 \ve^2\nnb \\
%---------------------
% term 9
\ar m_{\Lambda_b}^2  s \Big[(1 - r)^2  + 4 \sqrt{r} s - s^2 \Big] v^2  \ve
G_5 \ve^2 \nnb \\
%---------------------
% term 10
\ar 4 m_\ell^2  (1 - 2 \sqrt{r} + r - s) s \ve G_6 \ve^2 \nnb \\
%---------------------
% term 11
\ek 4 (1 + \sqrt{r}) (1 - 2 \sqrt{r} + r - s) (2 m_\ell^2  + m_{\Lambda_b}^2  s) 
\mbox{\rm Re}[F_1^\ast F_2] \nnb \\
%---------------------
% term 12
\ek 4 m_{\Lambda_b}^2  (1 + \sqrt{r}) (1 - 2 \sqrt{r} + r - s) s v^2  
\mbox{\rm Re}[F_4^\ast F_5] \nnb\\
%---------------------
% term 13
\ek 8 m_\ell^2  (1 - \sqrt{r}) (1 + 2 \sqrt{r} + r - s) 
 \mbox{\rm Re}[F_4^\ast F_6] \nnb\\   
%---------------------
% term 14
\ek 4 (1 - \sqrt{r}) (1 + 2 \sqrt{r} + r - s) (2 m_\ell^2  +
m_{\Lambda_b}^2  s) \mbox{\rm Re}[G_1^\ast G_2] \nnb\\   
%---------------------
% term 15
\ek 4 m_{\Lambda_b}^2  (1 - \sqrt{r}) (1 + 2 \sqrt{r} + r - s) s v^2  
\mbox{\rm Re}[G_4^\ast G_5] ] \nnb\\   
%---------------------
% term 16
\ar 8 m_\ell^2  (1 + \sqrt{r}) (1 - 2 \sqrt{r} + r - s) 
\mbox{\rm Re}[G_4^\ast G_6]
\Bigg\}~, \nnb
%------------------------
\eea
\bea
\Gamma_2 (s) \es
- 8 m_{\Lambda_b}^4  v^2 \lambda(1,r,s) \Big[
\ve F_1 \ve^2  + \ve F_4 \ve^2  + 
\ve G_1 \ve^2  + \ve G_4 \ve^2  - 
s \Big(\ve F_2 \ve^2 + \ve F_5 \ve^2  + 
\ve G_2 \ve^2  + \ve G_5 \ve^2 \Big)\Big]~. \nnb
\eea
%------------------------
%
%
The differential decay width for the $\Lambda_b^\ast \rar N \ell^+\ell^-$
transition cam be obtained from the differential decay width for the
$\Lambda_b \rar N \ell^+\ell^-$ by making the following replacements:
$F_i \rar \widetilde{G}_i$, $G_i \rar \widetilde{F}_i$, $ m_N \rar - m_N$,
and by changing the sign in front of the terms $\mbox{\rm Re}[F_4^\ast
F_5]$, $\mbox{\rm Re}[F_4^\ast F_6]$, and $\mbox{\rm Re}[G_4^\ast G_5]$,
as well as $m_{\Lambda_b} \rar m_{\Lambda_b^\ast}$, $s \rar
s^\prime = q^2/m_{\Lambda_b^\ast}^2$, and $r \rar r^\prime =
m_N^2/m_{\Lambda_b^\ast}^2$.

\newpage

\section*{Figure captions}
{\bf Fig. (1)} The dependence of the differential branching ratio 
for the $\Lambda_b \rar N \mu^+ \mu^-$ transition on $s$,
at $s_0=40~GeV^2$, and $M^2=25~GeV^2$. \\ \\
{\bf Fig. (2)} The same as in Fig. (1), but for the $\Lambda_b \rar N \tau^+
\tau^-$ transition. \\ \\
{\bf Fig. (3)} The same as in Fig. (1), but for $\Lambda_b^\ast \rar N \mu^+
\mu^-$ transition. \\ \\
{\bf Fig. (4)} The same as in Fig. (2), but for the $\Lambda_b^\ast \rar N \tau^+
\tau^-$ transition.

\newpage

\begin{figure}[t]
\vskip -1.5cm
\begin{center}
\scalebox{0.785}{\includegraphics{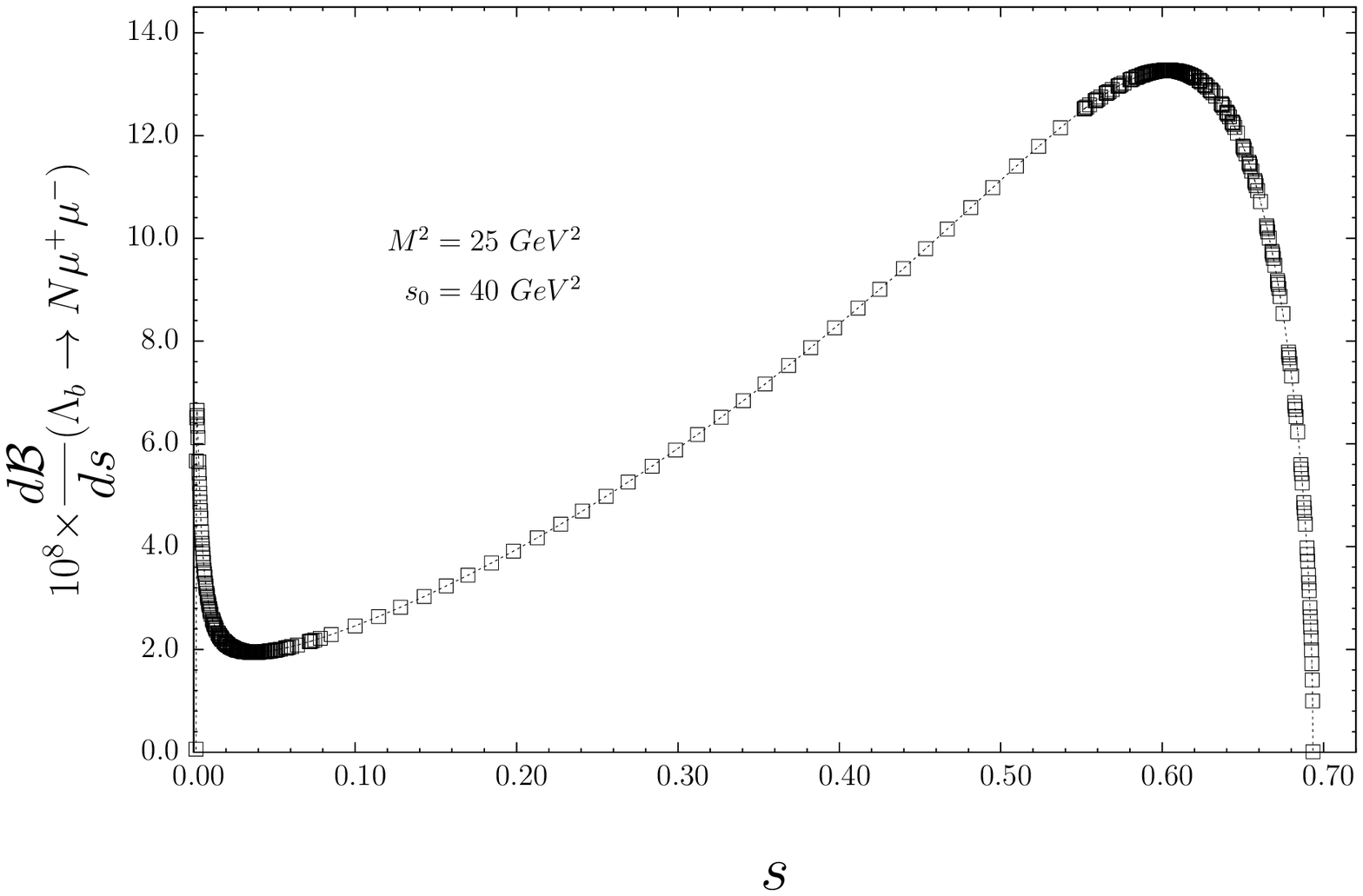}}
\end{center}
\vskip -12.0cm
\caption{}
\end{figure}

\begin{figure}[b]
\vskip -1.5cm
\begin{center}
\scalebox{0.785}{\includegraphics{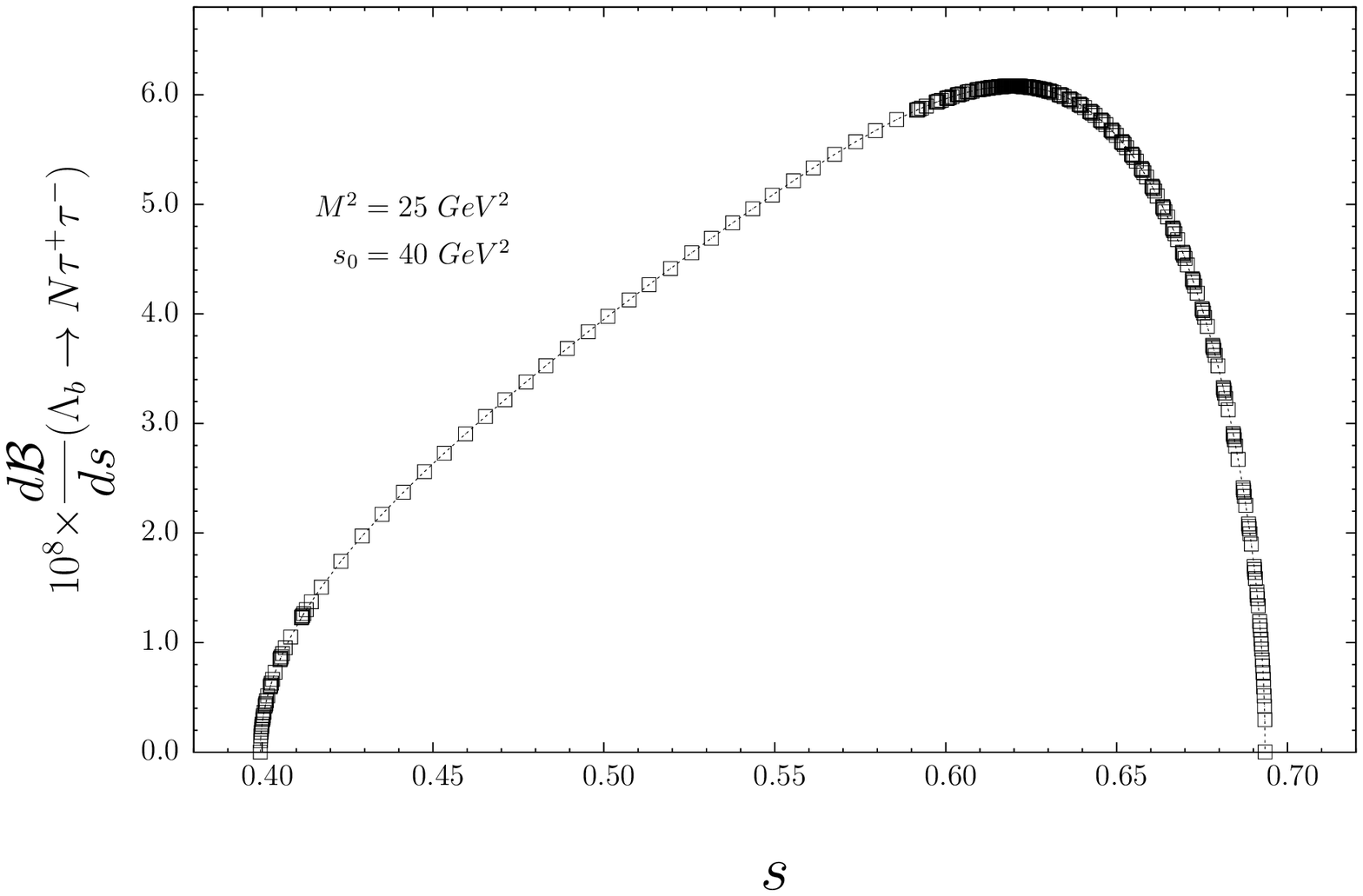}}
\end{center}
\vskip -12.0cm
\caption{}
\end{figure}

\begin{figure}[t]
\vskip -1.5cm
\begin{center}
\scalebox{0.785}{\includegraphics{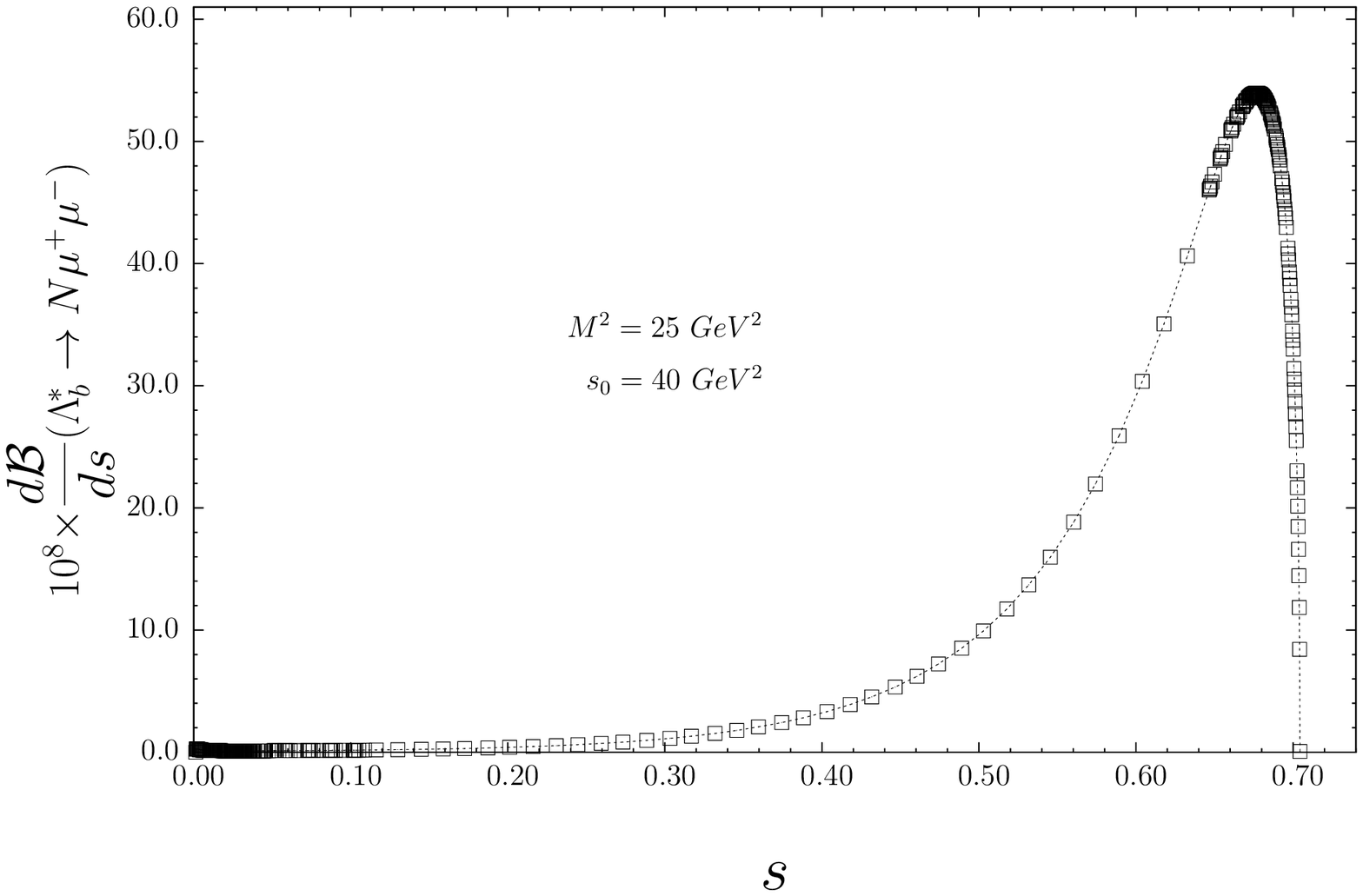}}
\end{center}
\vskip -12.0cm
\caption{}
\end{figure}

\begin{figure}[t]
\vskip -1.5cm
\begin{center}   
\scalebox{0.785}{\includegraphics{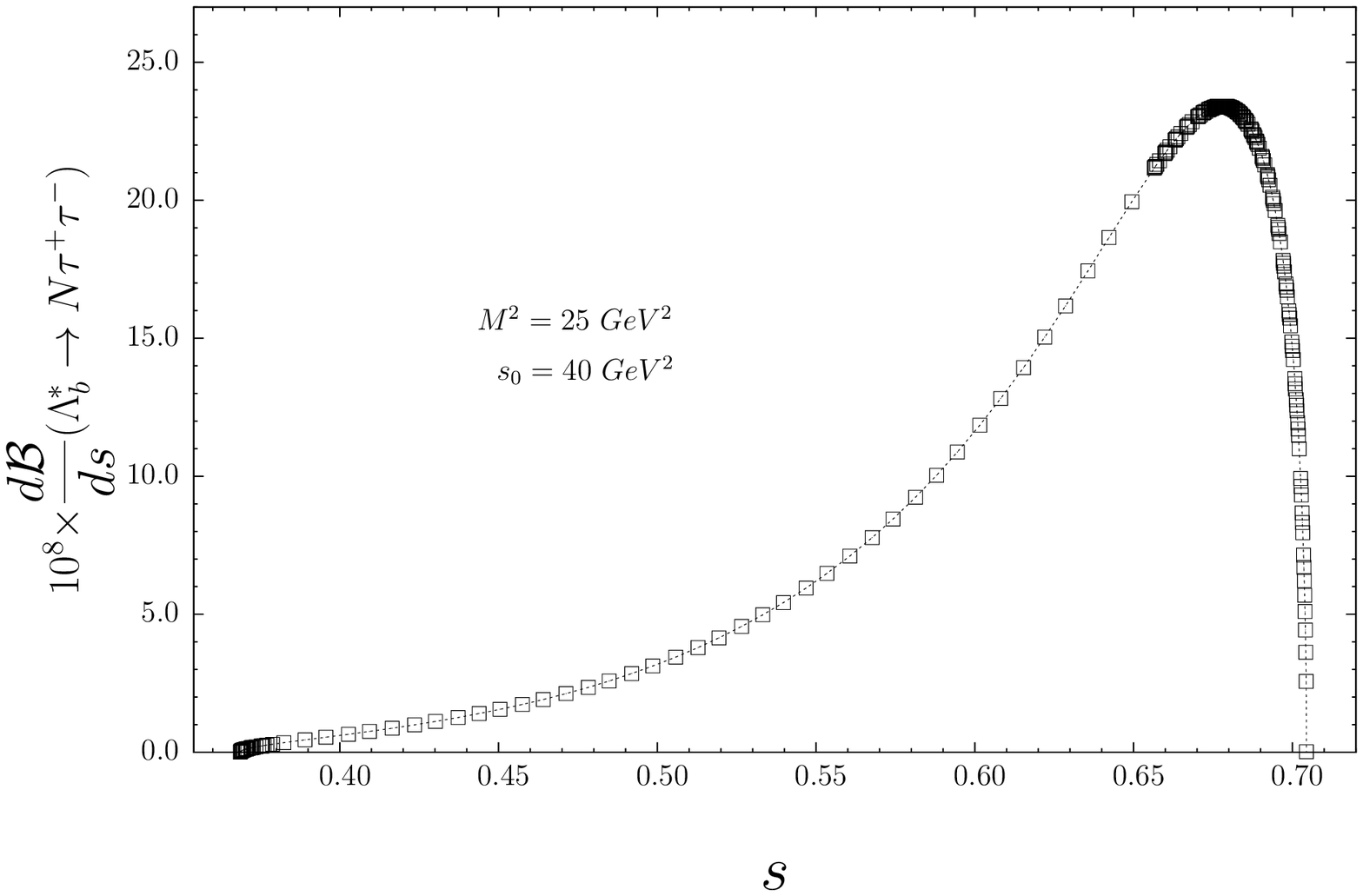}}
\end{center}  
\vskip -12.0cm                             
\caption{}  
\end{figure}

\end{document}